\documentstyle[a4,12pt]{article}
\begin{document}

\begin{center}

{\large\bf Multiplicity Distributions in QCD and $\lambda\varphi^3_6$-model}

\vspace{2mm}

I.M.~Dremin, V.A.~Nechitailo,

{\it P.N.~Lebedev Physical Institute, Moscow, Russia}

M.~Biyajima

{\it Shinshu University, Matsumoto, Japan}
\end{center}

\begin{abstract}
It is shown that QCD with vector gluons predicts
drastically different multiplicity distributions
compared to the scalar $\lambda\varphi^3$-model
(in 6 dimensions). In particular, the QCD predicted
minimum of cumulant moments of the distributions
at ranks about $5\div7$ does not reveal itself in
the scalar model. The oscillations at higher ranks
survive only. Thus, the experimental fact of that
minimum at ranks about $5\div7$ supports the vector
nature of gluons and singular kernels in QCD and
gluodynamics.
\end{abstract}

\section{Introduction}

The multiplicity distribution of particles (partons)
in QCD (or gluodynamics) has been a long-standing problem.
The solutions of equations for generating functions of
multiplicity distributions have been obtained in the lowest
double logarithmic approximation \cite{Dok}.
The shape of the distribution appears to be completely
different  from that observed in multiparticle production
experiments for secondary hadrons, even though the energy
dependence is rather reasonable and, moreover, the shape
fulfils the so-called KNO-scaling \cite{KNO}. The theoretical
shape is much wider. In terms of the moments of the distribution
it implies much faster increase of them with rank increasing,
compared  to experimental data.

Meantime, the theoretical attempts to take into account the
conservation laws more precisely were published \cite{GM,CT,D}.
The multiplicity distribution became rather narrow\cite{D}.
However, in view of some additional assumptions it is difficult to
estimate how rigorous the results are. Moreover, the theoretical
moments of the distributions differ from experimental ones.

The problem was solved when the solution of the equations for
generating functions was found at higher (than double-logarithmic one)
approximations in gluodynamics \cite{Dr,DN} and QCD \cite{DLN} in case of the
running coupling constant. Later, the exact solution of these equations at
fixed coupling constant was obtaned\cite{DH}.
The most remarkable prediction \cite{Dr} of these findings is the
minimum of the cumulant moments of the multiplicity distribution
at the values of ranks close to 5 and subsequent oscillations of those
moments at higher values of the rank \cite{DN} which have been
confirmed in experiment at analysis of hadron multiplicity
distributions in multiparticle production at high energies for
various colliding particles \cite{DAB} (see also the review paper\cite{DUFN}).

Let us note also that no phenomenological distributions well known
and widely used in physics (i.e. those distributions of probability
theory as Poisson, negative binomial, geometric, fixed multiplicity)
can reproduce such a behaviour. In particular, the negative binomial
distribution gives rise to the positive and monotonically decreasing
ratio of cumulant to factorial moments, while QCD predicts the negative
values of this ratio at the first minimum and its subsequent oscillations
about the abscissa (where the ranks of the moments are plotted).
Even the so-called modified negative binomial distribution, specially
designed empirically \cite{ChTch} to get the best fit of experimental
multiplicity distributions has failed in some cases \cite{BNS} to produce
the relevant minima and oscillations of the moments. Thus the analysis
of multiplicity distributions in terms of cumulant and factorial moments
as well as of their ratio is at present the most sensitive method of
revealing the specific typical details of these distributions.

The very existence of the minimum in QCD and its location are closely
defined by the singularity of the gluon kernel (the vector nature of
gluons) and by the value of the coupling constant in QCD.
In that connection it is of interest to understand how important
the both factors are. According to the  formulae of the paper \cite{Dr},
the main contribution to the minimum location is inverse proportional
 to the QCD anomalous dimension (i.e. to the coupling constant)
and therefore varies in QCD in the limits of its "running" property
i.e. comparatively slow and hard to observe if, especially, one
takes into account that the ranks of the moments can be integer only.
The more important property is the vector nature of massless gluons
giving rise to the singularity of the kernel of the equation for the
generating function. That is why it would be desirable to confront
the  QCD (or gluodynamics) predictions to those of the scalar fields
model. Fortunately, there is such a model\cite{Hwa} possessing, besides
others, the property of the asymptotical freedom what helps to reduce
the impact (even if it is not very essential) of the problem of the
value of the coupling constant. This is the $\lambda\varphi^3$-model
in  the 6-dimensional space-time.

Therefore the purpose of the present paper is to get the
knowledge of the behaviour of the multiplisity distribution
moments for the $\lambda\varphi^3$-model in  the 6-dimensional space-time
and to compare the obtained results with those of QCD and of
the simplest phenomenological distributions of the probability
theory. The main conclusion which we have got from it
is that the vector nature of gluons is very crucial for the
shape of the multiplicity distribution and qualitatively changes
its moments behaviour so that QCD confirms its predictive power
once again even at purely partonic level while the scalar model
has been unable to reproduce the qualitative features of experimental
data.

\section{Equations and their solutions}

The theoretical problem, we are addressing at, is as follows. Let a
scalar, strongly virtual particle (with high time-like 4-momentum
squared) has been produced in some collision. During its evolution
the virtuality decreases due to emission of other scalar particles.
What is the multiplicity distribution of secondary particles if the
emission is controled by the $\lambda\varphi^3$-interaction in the
6-dimensional space-time? This problem is analogous to the production
and evolution of the pair of strongly virtual quark and antiquark
in $e^{+}e^{-}$-annihilation. The increased dimensionality of the
space-time, as we have mentioned, is necessary here just to get the
asymptotical freedom in that theory and to make it closer to the real
situation in QCD (or in gluodynamics).

The evolution of such a "scalar jet" is described by the equation for
generating functions (or functionals) analogous to the commonly used
"birth and death equation" and different from QCD equations by its
kernel, describing the interaction vertex in the theory.
However, just this distinction strongly influences the multiplicity
distribution  and its moments. The principal difference is that the
QCD kernel is singular while in the scalar model it is regular.
Physics corollary is an approximate equipartition of the "parent"
energy among its "children" in the scalar model while in QCD the
energy is shared in unequal parts i.e. one of the produced partons
used to get much higher energy than another one.

Let us  turn to our problem and remine some general notations.
The generating function is

\begin{equation}
G(y,z) = \sum^{\infty}_{n=0} (1+z)^n P_n(y),
\end{equation}
where $P_n(y)$ is the multiplicity distribution,
$y=\ln(Q/Q_0), Q$ is the jet virtuality, $Q_0=$const.

The normalized factorial moments ($F_q$) and cumulants ($K_q$)
are defined by the generating function as

\begin{equation}
G(z)=\sum_{q=0}^{\infty}\frac{z^{q}}{q!}\langle n\rangle^{q}
F_{q}, \qquad \qquad
\ln G(z)=\sum_{q=0}^{\infty}\frac{z^{q}}{q!}\langle n\rangle^{q}
K_{q},
\label{FP}
\end{equation}
where $\langle n\rangle$ is the average multiplicity of particles (partons).

The generating function satisfies the non-linear integro-differential
equation \cite{Dok} (the prime denotes the $y$-derivative):

\begin{equation}
G^{\prime}(y)=\int_{0}^{1}d\xi K(\xi)\gamma_{0}^{2}
[G(y+\ln \xi)G(y+\ln (1-\xi))-G(y)] ,
\label{AP}
\end{equation}
where $K(\xi)$ is the kernel of the equation and in our cases is written
as: \\
1)~for gluodynamics \cite{Dok}:
\begin{equation}
K(\xi)=1/\xi-(1-\xi)[2-\xi(1-\xi)] \label{kerG},
\end{equation}
2)~for  $\lambda \varphi^3$-model in 6 dimensions \cite{Hwa}:
\begin{equation}
K(\xi)=6\xi(1-\xi). \label{kerPhi}
\end{equation}
$\gamma_0^2=2N_c \alpha_s/\pi,\ \alpha_s=2\pi/11y$ in QCD.
In $\lambda \varphi^3$-theory one does not have any physics
normalization and has to rely on QCD $\gamma_0^2$
in choosing the numerical value of $\lambda$.

Now we discuss $\lambda \varphi^3$-model directly and consider
first the running coupling case, following the method of the approximate
solution used in  \cite{Dr},\cite{DN} for gluodynamics. After the Taylor
series expansion of the generating function at point $y$ has been done,
one gets

\begin{equation}
G^{\prime}(y)=\gamma_0^2\{ G(y)(G(y)-1)+
G(y)\sum^{\infty}_{n=1} (-1)^n h_n G^{(n)}(y) +
\sum^{\infty}_{n,m=1} (-1)^{n+m} h_{nm} G^{(n)}(y) G^{(m)}(y)\},
\end{equation}
where\footnote{We are using the same notations $h_i$ for the coefficients in
the $\lambda\varphi^3$-model as in gluodynamics \cite{Dr},\cite{DN},
and hope that it does not produce any misunderstanding even though their
numerical values differ in the both cases.}
$$h_n = \frac{12}{n!}(-1)^n \int^{1}_{0}d\xi \xi(1-\xi) \ln^n
(1-\xi) =12(\frac{1}{2^{n+1}}-\frac{1}{3^{n+1}} ) ,
$$
$$ h_{nm} =
\frac{6}{n!m!}(-1)^{n+m} \int^{1}_{0}d\xi \xi(1-\xi)\ln^n(\xi)\ln^m (1-\xi)
$$
or within the approximation used in \cite{DN},
one gets

\begin{equation}
(\ln G(y))^{\prime}=\gamma_0^2\{ G(y)-1
- h_1 G^{\prime}(y) + h_2 G^{\prime\prime}(y)
+ h_{11} G^{\prime}(y)(\ln G(y))^{\prime} \},
\label{lnG}
\end{equation}
where  $ h_1 = 5/3,\quad h_2 = 19/18,\quad h_{11}=(37-3\pi^2)/18$.

The anomalous dimention $\gamma$ of the $\lambda\varphi^3$-model is defined similar
to QCD :
$\langle n\rangle = \exp(\int^y \gamma(y^{\prime})dy^{\prime})$ and we
substitute (\ref{FP}) to (\ref{lnG}).
The coefficients in front of $z^q$-terms should be equal on both
sides wherefrom the moments satisfy the equation

\begin{equation}
k_q=\frac{1}{1-H_q^0}
\sum^{q-1}_{l=1}k_{q-l}f_l \left\{ \frac{H^0_q}{l} +
\gamma^2_0 x[\frac{h_2}{l}+\frac{h_{11}}{q}]\right\},
\label{kq}
\end{equation}
$ x\equiv q\gamma,
H^0_q \equiv \gamma^2_0[1/x - 5/3+h_2\gamma^{\prime}/\gamma],
k_q\equiv K_q/(q-1)!, f_q\equiv F_q/(q-1)!$
and we have used the relation

\begin{equation}
f_q = k_q + \sum^{q-1}_{l=1}\frac{k_{q-l}f_l}{l}.
\label{iden}
\end{equation}
The recurrent formula (\ref{kq}) has been used to calculate $k_q$
and to get $f_q$ with the help of eq.(\ref{iden})
as well as their ratio
$H_q$ if $\gamma$ is known.

The equation for $\gamma$ obtained from (\ref{kq}) in the similar way
to  (\ref{lnG}) but for $q=1$ is written as

\begin{equation}
\gamma - \gamma^2_0 [1-\frac{5}{3}\gamma + h_2
(\gamma^{\prime}+\gamma^2)]=0 ,
\end{equation}
what represents $\gamma$ in terms of $\gamma_0$ as
$$ \gamma=\gamma^2_0[1-\frac{5}{3}\gamma^2_0]+O(\gamma^6_0).$$

The results of computing $H_q$ according to (\ref{kq})
for $\gamma_0=0.48$ are shown in Fig.1. In distinction to QCD results
obtained within the same approximation in $q\gamma$,
there are still no oscillations and $H_q$ tends to  constant
asymptotics (it reminds of QCD in the lower order in $q\gamma$ -- see
\cite{Dr}).

Varying $\gamma_0$ within the wide limits, one does not observe any essential
qualitative changes. The minimum is slightly shifted to larger values of $q$
at smaller $\gamma_0$.

Thus we notice that the absence of the singular term in the kernel
(\ref{kerPhi}) in the $\lambda\varphi^3$-model compared to the gluodynamics
(the formula (\ref{kerG})) gives rise to qualitatively different behaviour
of the moments.

In the case of the fixed coupling constant, the equation (\ref{AP})
can be solved exactly \cite{DH}. Assuming that the $y$-dependence
is contained completely in the average multiplicity $\langle n \rangle$
and is of the kind $\langle n \rangle = \exp(\gamma y)$,
one gets the recurrent formula for the factorial moments

\begin{equation}
\bar{f}_q =
\frac{6\gamma^2_0}{x(2+x)(3+x)+\gamma^2_0(x-1)(x+6)}
\sum^{q-1}_{l=1}\bar{f}_{q-l}\bar{f}_l,
\label{fq}
\end{equation}
where $\bar{f}_q=f_q \Gamma(x+2)/q=F_q\Gamma(x+2)/q!$
and
$$H_q=1-\sum^{q-1}_{l=1}\frac{H_{q-l}}{l}\frac{f_{q-l}f_l}{f_q}.$$

The analogous formula in gluodynamics looks like
(it can be easily obtained from the formulae of \cite{DH}
if the quark degrees of freedom are neglected)

\begin{equation}
\bar{f}_q = \frac{\gamma^2_0}{x-\gamma^2_0 M_q}
\sum^{q-1}_{l=1}N_{q,l}\bar{f}_{q-l}\bar{f}_l,
\label{fqDH}
\end{equation}
where
$$ M_q=\frac{1}{x}+\Psi(1)-\Psi(1+x)+\frac{11}{12}-
\frac{2}{1+x}+\frac{1}{(2+x)(3+x)}
$$
$$
N_{q,l} = \left(\frac{1+x}{l\gamma}-1\right)
          \frac{1}{(1+l\gamma)(1+(q-l)\gamma)} +
         \frac{1}{2(2+x)(3+x)}.
$$

In Figs.\ 2 and 3 we show the results of computing according to
formulae (\ref{fq}) and (\ref{fqDH}) correspondingly. One can conclude
from them that the influence of the singularity in the kernel
$K(\xi)$ on the location of the minimum consists in its shift
to smaller values of ranks $q$ ( from $q_1 \approx 14$ to
$q_2\approx q_1/2=7$) with the accompanying diminishing of the
"period" of the oscillations.

\section{Conclusion}
The main conclusion one gets from above consideration is that
the scalar $\lambda\varphi^3$-model in 6 dimensions with approximate
equipartition of energy among the produced particles gives rise
to the qualitatively different predictions about the behaviour of
the moments of multiplicity distributions compared to the gluodynamics
or/and QCD, where the vector nature of massless gluons implies the
singularity of the kernel of the equation and, therefore, the
drastically unequal shares of energy for produced particles.
This fact provides the minimum in the ratio of cumulants to
factorial moments at ranks near $5\div7$ in gluodynamics and QCD,
while in the scalar model such a minimum does not appear and the
oscillations at higher values of ranks survive only.

From the purely theoretical approach, this conclusion is hardly
extremely important. However, it becomes essential when one confronts
it to the experimentally known facts about hadron
multiplicity distributions in real events \cite{DAB,DUFN}. It happened that
the minimum of $H_q$ at $q\sim5$ was observed \cite{DAB} in experimental
data just after its prediction \cite{Dr} in gluodynamics.
Even though this prediction was done about partons and not about final
hadrons, the very fact of the presence of the minimum with the same
location as predicted looks very impressive. In combination
with the inability of the phenomenological models as well as of the
theory field model  considered above to reproduce the existence of
the minimum and its location, it confirms once again our belief in the
predictive power of QCD. Other numerous facts (the hump-backed plateau
of the rapidity distribution, the heavy quark effects etc. -- see in
detail in \cite{Dok} and in later papers) support the conclusion that the
qualitative effects of QCD predicted at the parton level find out
their confirmation by the experimental data, and Monte-Carlo simulations
provide the quantitative estimates of the influence of hadronization.

\bigskip
\noindent {\Large\bf Acknowledgements}
\bigskip

This work has been supported by Russian Fund for Fundamental Research
(grant 93-02-3815), by INTAS (grant 93-79) and by JSPS.
MB is partially supported by the Japanese Grant-in-aid
for Scientific Reserach from the Ministry of Education, Science and
Culture (No. 06640383).

\end{document}